\documentclass[12pt]{iopart}
\usepackage{graphicx}
\usepackage{url}

\usepackage{xcolor}
\usepackage{ulem}

\begin{document}

\title[Analysis of optical pattern formation on glass]{Analysis of optical pattern formation on glass: exploring light phenomena in the International Physicists’ Tournament }

\author{Maressa P Sampaio$^1$$^*$, Renan G Alvim$^1$, Felipe K Kalil$^1$, Maria C O Aguiar$^1$, Ubirajara Agero$^1$}

\address{$^1$ Departamento de Física, Universidade Federal de Minas Gerais, C.P. 702, 30123-970 Belo Horizonte, Minas Gerais, Brazil}

\ead{$^*$ mps2020@ufmg.br}

\begin{abstract}
    Motivated by a problem from the 2023 International Physicists' Tournament, we investigate the formation of particular patterns when light passes through glass. Experimentally, we use various glass plates, registering each reflected and transmitted outcome. Thus, we find the condition to form a common pattern, namely, randomly scratched plates produce the halos. We engineer other optical patterns by means of the specular holography method. Using the Defocusing Microscopy technique, we study the geometric properties of the glass, establishing the physical and mathematical model for a simulation. Our computational results agree well with the experimental ones, leading us to conclude that the pattern formation is governed by specular reflection on various points along the scratches. Our findings give an initial explanation for the pattern formation on glass, in particular for the halo design. We also show that the IPT open questions help students learn physics concepts and develop research skills.

\end{abstract}

\vspace{2pc}

\maketitle

\section{Introduction}
In optics, the interaction of light with rough surfaces is fundamental to understanding everyday effects, considering that objects are not perfectly smooth. For instance, the light scattering phenomenon involves the redirection of light by oscillating charges, and it approaches the limit of geometric optics for large particles in comparison to the wavelength of light \cite{hahn2009light}. The intensity and profile of the scattered light are studied using complex methods and approximations, as summarized in Refs.~\cite{elfouhaily2004critical, sylvain2005diffuse, maradudin2010light,  
schroder2011modeling,
pinel2013electromagnetic}. We can divide these methods into two main categories: (i) in analytical theories \cite{rice1951reflection, beckmann1987scattering}, certain approximations are used to solve Maxwell's equations, which have limitations and a domain of validity, while (ii) in numerical models \cite{warnick2001numerical, saillard2001rigorous} Maxwell's equations are solved in differential or integral form. Various parameters and regimes can be analyzed: small to large roughness, different scattering materials, and roughness distributions. Thus, results are usually reported based on specific cases, both for analytical \cite{ davies1954reflection, bennett1961relation,  croce1976light, ma2022implementation} and numerical \cite{toporkov2000numerical, yan2022deep, zhang2025generalized} studies.

These mentioned works concentrated on developing broad theories for the interaction between light and surface roughness, not focusing on the formation of optical patterns. Interestingly, surface roughness can be used to create holograms. Experiments with plastic paper and a compass revealed a technique, called specular holography, to create holograms by hand \cite{inproceedings, duke2013drawing}. Arched scratches on the surface act as a curved mirror and form a tridimensional (3D) image. A computational method successfully predicted the bright spots that make up the hologram by applying geometric optics and specular reflection \cite{li2017general}, which greatly simplifies the light-surface interaction in this scope. Thus, the relation of roughness and pattern formation started to be explored in the holographic technique.

Even so, the unguided formation of certain patterns on the surface of reflective materials, mainly glass, remains to be described. In this paper, we start from the problem ``Glass halo", proposed for the International Physicists' Tournament (IPT) 2023 \cite{ipt}. The IPT is the largest physics competition between teams of university students from different countries. Each year, 17 open problems are proposed, and the teams have a couple of months to prepare their solutions. During the tournament rounds, a team presents and debates their solution with other teams \cite{iptrules}. These open problems provide a research prospect to undergraduate students, allowing them to creatively elaborate experiments and apply physics concepts while working as a team. Thus, when working on IPT problems, students develop relevant abilities to their academic career, such as research planning and effective communication.  In particular, we worked on this problem while preparing for the Brazilian Physicists' Tournament (BPT), which follows the same format and selects the Brazilian team that will participate in the next IPT. During this research process, we explored various possibilities and debated with our colleagues and professors until reaching the solution detailed in this paper. Our results were presented in the 2022 BPT and the 2023 IPT Conference; for the first time the conference abstracts were published as proceedings \cite{sampaio_2024_10899678}.

\begin{figure}
    \centering
    \includegraphics[scale=0.29]{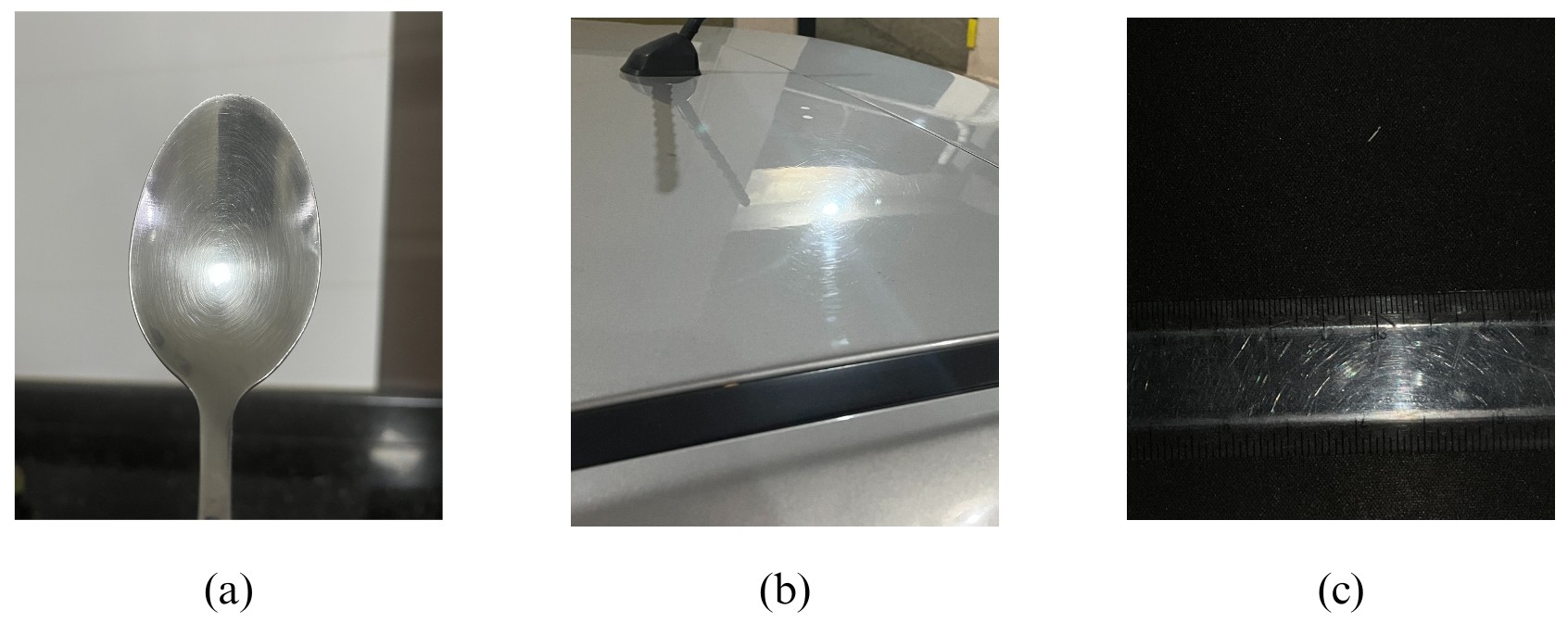}
    \caption{Halo pattern seen on (a) a spoon, (b) a car surface and (c) a ruler.}
    \label{fig introducao}
\end{figure}

The ``Glass halo" problem statement goes as: ``Glittering circles can be seen when light from a source with small angular size passes through a glass. On closer examination they appear to be composed of small scratches and structural inhomogeneities. In some cases, specific rays can be seen, diverging from the light source. Under which conditions such circle halos and lines can be seen? Investigate their geometrical properties and what shapes you can engineer."
Accordingly, we started by experimenting with microscope glass plates and registering the pattern outcome. A geometric investigation laid the basis for a simulation that agrees with the geometric optics simplification proposed by \cite{li2017general}. 
As we will show, the resulting optical design allows us to specify the roughness pattern in various materials. This identification is an initial progress when dealing with everyday objects that do not always appear rough to the human eye; for instance, the patterns discussed here can appear in glass windows, cellphone screens and many other reflective surfaces, as seen in Fig. \ref{fig introducao}. From the viewpoint of applications, the understanding of pattern formation associated with roughness improves studies of surface polishing. 

Our goal is (i) to present the steps taken to arrive at our solution for this IPT problem and (ii) describe the benefit of these open questions in teaching physics concepts and scientific methods. For example, while tackling the Glass Halo problem, we planned experiments and pursued an inquiry-based learning, since we could not find an explanation for it in the literature and had to test various scenarios to find the right conditions. Additionally, our solution can now help undergraduate students visualize optical concepts with an accessible experiment and learn geometric optics simulation. 

The remaining sections are organized as follows. In Sec. \ref{methods}, we present all methods: experimental procedures with glass plates, simulation, and experimental process with the specular holography technique. In Sec. \ref{results}, we present the experimental results, comparing them to the simulation when suitable; we also discuss the points for additional studies along with the pedagogical application of our results. Finally, our concluding remarks are stated in section \ref{conclusao}.

\section{Methods}\label{methods}

\subsection{Experiments}\label{experiments}

\begin{figure}
    \centering
    \includegraphics[scale=0.28]{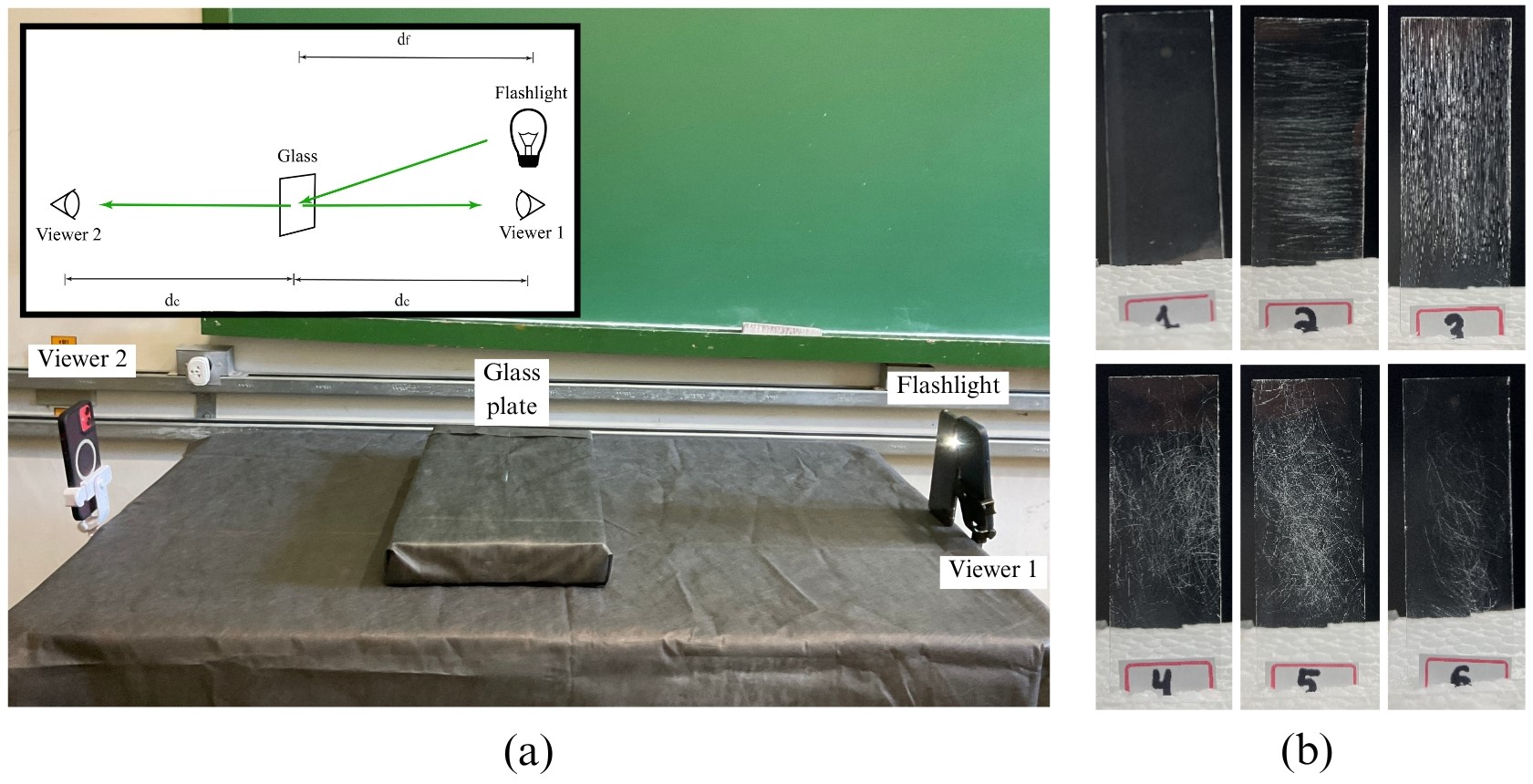}
    \caption{(a) Experimental setup. Inset: Sketch showing the important parameters. For illustrative purposes only. (b) Glass plates described in Table \ref{Table 1}. Since light is not hitting the glass surface perpendicularly (it is coming from above), we do not see the formation of optical patterns here.}
    \label{Fig. 1}
\end{figure}

We started our investigation with experiments varying one parameter at a time: (a) first the glass surface and (b) then the distance between the glass plate and the camera-light source combination:  see Fig.~\ref{Fig. 1}a for our experimental setup. We used polished as well as mechanically scratched glass plates. The materials and their specifications are listed in Table \ref{Table 1}. The cellphone flashlight is composed of light emitting diodes (LEDs) and its intensity was constant during all the experiments carried out for this paper. Also, the glass surface is always parallel to the cellphones.

\begin{table*}[b]
    \renewcommand{\arraystretch}{1.3}
    \centering
    \begin{tabular}{c c c}
    \hline
     Item   & Characterization \\
    \hline
    Plate 1  &  Polished glass plate\\
    \hline
    Plate 2 &  Horizontally scratched plate with a 150 grit sandpaper \\
    \hline
    Plate 3 & Vertically scratched plate with a 150 grit sandpaper \\
    \hline
    Plate 4 & Randomly scratched plate with a 80 grit sandpaper\\
    \hline
    Plate 5 & Randomly scratched plate with a 150 grit sandpaper\\
    \hline
    Plate 6 & Randomly scratched plate with a 280 grit sandpaper\\
    \hline
    Flashlight & Cellphone flashlight with constant intensity\\
    \hline
    Measuring tape & Milimiter tape attached to the table\\
    \hline
    Cellphone camera & 12MP Dual camera with digital zoom up to 5x\\
    \hline
         
    \end{tabular}
    \caption{Materials used in the experiments with glass. All plates have the same dimensions: $a_1=(7.6 \pm 0.1)$\,cm, $a_2=(2.5 \pm 0.1)$\,cm, and $a_3=(0.10 \pm 0.05)\,$cm, where $a_1$ is the height, $a_2$ is the width, and $a_3$ is the thickness.}
    \label{Table 1}
\end{table*}

During the first part, we took photos of the reflected (viewer 1) and transmitted (viewer 2) patterns for each glass plate (see again Fig.~\ref{Fig. 1}). In all these cases, the distance between the glass surface and the flashlight was always $d_f=(0.55\pm0.01)\,$m. Similarly, the distance between the surface and both cameras (observers) was always $d_c=(0.55\pm0.01)\,$m to avoid obstructions and shadows in the way. 

In the second round of experiments, we varied the distances $d_f$ and $d_c$ for plates 5 and 6, taking photos of the reflected and transmitted patterns each time. Starting at $d_f=d_c=(0.55\pm0.01)\,$m, we reduced ten centimeters at a time until $d_f=d_c=(0.25\pm0.01)\,$m.

\subsection{Scratch reconstruction}\label{reconstrução_methods}

Subsequently, we studied the geometrical properties of the glass plates in order to characterize the scratches properties. In this context, the Defocusing Microscopy (MD) technique \cite{ agero2004defocusing} is used to reconstruct the thickness and curvature profiles of phase objects, such as red blood cells \cite{roma2014total} and waveguides \cite{Lages:18}. The basic idea behind this technique is that phase objects can be observed in bright-field microscopes when defocused.

That means this method could be used to reconstruct the profile of the
scratches on selected regions of plates 4, 5, and 6, since they are phase objects, which otherwise would be transparent under bright-field microscopes. In MD, the image contrast for the scratch is given by
\begin{equation}
C(z_{f1}) = -\frac{I - I_0}{I_0}= \frac{\Delta n}{n_o} \, z_{f1}\, \nabla^2 h,
\end{equation}
$I$ being the intensity of the image, $I_0$ the mean intensity  of the image, $\Delta n$ being the difference between the refractive index of air and the plate glass, $n_o$ the refractive index of the objective immersion fluid (in our case, $n_0=1$ since we use a air objective), $z_{f1}$ the distance to a reference plane and $h$ the object profile. The equation show us that the contrast of these images is proportional to the Laplacian of the objects surface, in other words, to the local curvature of the sample.

The technique consists in selecting an area of interest of the plate and adjusting the microscope to obtain an image with the least possible contrast as a reference plane $z_{f0}$.  After that, two images are captured, one defocused above $(z_{f1})$ and the other below $(z_{f2})$. These images, each having different contrasts $C(z_{f1})$ and $C(z_{f2})$, can be subtracted from one another, pixel by pixel, to create a third image representing the difference of contrast between the two, described by:
\begin{equation}
    C(z_{f2}) - C(z_{f1}) = \Delta n \, (z_{f2}-z_{f1}) \, \nabla^2 h,
\end{equation}
Since we want the scratch profile $h$, we have:
\begin{equation}
     \nabla^2 h = \frac{C(z_{f2}) - C(z_{f1})}{\Delta n \, (z_{f2}-z_{f1})},
\end{equation}
This is a Poisson equation that can be solved in the Fourier space and the solution is given by

\begin{equation}
\Delta h (x,y) = \frac{p^2}{\Delta n (z_{f2} - z_{f1})N^2}  \mathcal{F}^{-1} \left\{ \frac{\mathcal{F}\{C(z_{f2}) - C(z_{f1})\}}{q \,^2(x) +q \,^2(y)} \right\},
\end{equation}
where $\mathcal{F}$ is the Fourier transform, $q(x)$ and $q(y)$ are the spatial frequencies of the image, $N$ the size of the image window and $p$ the size of each pixel in micrometers.

The scratched plates were analyzed in an inverted microscope Nikon Eclipse TiE using $10X$ (NA 0.55) and $40X$ (NA 0.25) objectives and a Silicon Video SV643M CMOS camera. We can estimate the lateral resolution of our images. For the 40x(NA 0.55) objective, the resolution is around $0.75\,\mu m$ and, for the 10x (NA 0.25) one, the resolution is around $1.10\, \mu m$. As for the measurements of the scratches depth in the direction of the light propagation, we have an increased resolution, since defocusing microscopy measures the phase difference of the light and is an interferometric technique, similar to other optical phase microscopy techniques. To estimate the sensitivity to measure the depth of the scratches, we can follow equation 10 in \cite{roma2014total}, which gives for a scratch with $10\,\mu m$ radius a $100\, nm$ sensitivity.

\subsection{Simulation}\label{sec_simulation}

Following Ref. \cite{li2017general}, our simulation recreates the reflection of light on scratched glass based on two main assumptions:

\begin{itemize}
    \item [-] in our experiments, the reflected and transmitted outcomes are equal;
    \item [-]the physical regime is that of geometric optics. 
\end{itemize}

\begin{figure}
    \centering
    \includegraphics[scale=0.6]{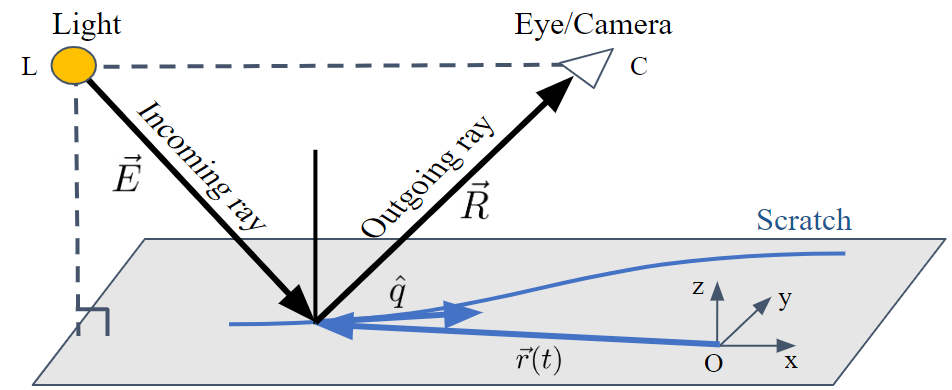}
    \caption{Simulation outline. The plane containing the light source and the detector is parallel to the plate, as seen in Fig. \ref{Fig. 1}; here the image is rotated to visualize better. The incoming ray reaches the scratch and is reflected towards the camera. The reflection is seen (plotted) if it obeys the tolerance criterion.}
    \label{Fig. 2}
\end{figure}

The first one is correct for a thin plate or when the observer is far from the surface, which are two conditions satisfied in our configuration - see \cite{li2017general} for a detailed discussion. Although the transmission and reflection coefficients are distinct and the intensity will differ, the pattern for both observers is the same. The second premise is a simplification that we will argue to be valid for this phenomenon based on (i) the good agreement between our experiments and simulations and (ii) the investigation of scratch size using defocusing microscopy. Consequently, we only analyze the reflected pattern and use the law of reflection, given by

\begin{equation}
    \hat{E}(t)\cdot\hat{q}(t)=\hat{R}(t)\cdot\hat{q}(t), \label{lawreflection}
\end{equation}
to plot the light rays. In equation \ref{lawreflection}, $\hat{E}(t)$ is the unit vector in the direction of the incoming ray, that is, corresponding to the vector that goes from the light source to a given point on the scratch, while $\hat{R}(t)$ defines the unit vector in the direction of the outgoing ray, i.e., the direction of the vector that goes from the point on the scratch to the camera (see Fig.~\ref{Fig. 2}). The unit vector $\hat{q}(t)$ defines the tangent line to the scratch curve. 
For a generic scratch curve, equation \ref{lawreflection} is not able to be solved analytically, and therefore needs root finding algorithms that naturally introduce a narrow tolerance for the equation. Numerically, what the computer does is finding values of $t$ that satisfy:

\begin{equation}
    \left|\hat{E}(t)\cdot\hat{q}(t)-\hat{R}(t)\cdot\hat{q}(t)\right|\leq TOL, \label{unfinishedTolerance}
\end{equation}
and since $|\hat{E}(t)|=1$, $|\hat{R}(t)|=1$ and $|\hat{q}(t)|=1$, the camera or our eyes will capture the reflected beam if the following relation is satisfied

\begin{equation}\label{cosine_diff}
    cos(\theta_1)-cos(\theta_2)\leq TOL. \label{cosdif}
\end{equation}

Also, scratches on everyday objects are random unless they undergo an ordered process. We assume that the scratches behave like a gap in the glass with semicircular cross section and are locally linear. This shape produces the following feature: the reflected ray can exit at any angle in the plane passing through the section, since it depends on the point that the light hits the semicircle. This implies that a scratch can reflect the ray at any angle in a plane passing through the point, which is a reasonable assumption, considering that we do not know the profile of the scrape. Lastly, the light source is considered point-like. See \ref{appendix_simulation} for a detailed explanation of these assumptions and an example.

In our simulation, written in Python and available at \cite{sampaio_2024_14718426}, first we generate the scratches (random or ordered) and then test the law of reflection for every point along the scratches. Only points that obey the cosine difference of Eq.~\ref{cosine_diff} for a given tolerance ($TOL=0.0125$) are visible in the plotted results.

We set the simulation parameters to match our experimental arrangement. Initially, the light source and the camera are located at $\vec{r}=(0.55 \mbox{ m}) \hat{z}$ (in the coordinate system defined in Fig.~\ref{Fig. 2}, for which $\hat{x}$ and $\hat{y}$ are in the plate plane), corresponding to the distances $d_f=d_c=(0.55\pm0.01)\,$m. At this configuration, we simulated the outcome for the scratched plates, varying the number of scratches on the plate for the random case.
Lastly, with a defined number of random scratches, we moved the camera and light positions in the simulation ten centimeters at a time until reaching $d_f=d_c=(0.25\pm0.01)\,$m as in the experiments.

Even though our simulation was based on a holography paper \cite{li2017general}, it is still valid, considering that the study is based on simple reflection and refraction laws, and we generate linear scratches to match the experiments, deferring from the reference, which uses scratches curved in a specific manner to form holograms - see Sec. \ref{holografia}. 

\subsection{Specular Holography}\label{holografia}

Besides the pattern formation of each glass plate, we relied on the specular holography technique \cite{inproceedings, duke2013drawing} to engineer new shapes. Although this technique can be applied to glass, we used plastic surfaces (CD cases; usually made of polystyrene \cite{neoonlineWhatPlastic}) since it was easier to manipulate and scratch. The materials for this experimental part are recorded in Table \ref{Table 2}.

\begin{table*}[hb]
    \renewcommand{\arraystretch}{1.3}
    \centering
    \begin{tabular}{c c c}
    \hline
     Item   & Purpose \\
    \hline
    Paper &  Used to draw the guiding blueprint \\
    \hline
    School compass &  Used to scrape the plastic\\
    \hline
    Jewel CD cases &  Scratching surface\\
    \hline
    Cellphone camera & 12MP Dual camera with digital zoom up to 5x\\
    \hline
         
    \end{tabular}
    \caption{Materials used while applying the specular holography technique.}
    \label{Table 2}
\end{table*}

\begin{figure}[hb]
    \centering
    \includegraphics[scale=0.5]{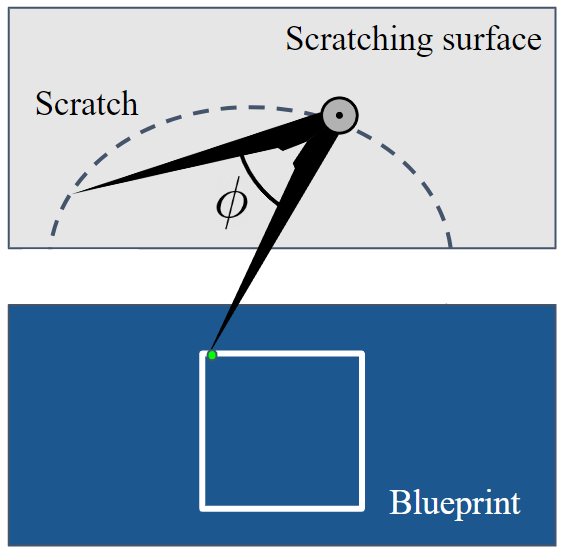}
    \caption{Schematic figure of the drawing process. One leg of the compass stays on the blueprint, following the points along the desired shape.}
    \label{Fig. 3}
\end{figure}

We start by drawing the desired shape on the paper that acts as the blueprint. Then, the fixed portion of the compass is placed on the blueprint, and we rotate the compass gently scratching the plastic surface with its metal leg (see Fig.~\ref{Fig. 3}). We apply this procedure to various points along the blueprint, always marking the plastic in a circular motion. The compass aperture angle was constant at $\phi=(35 \pm 1)\,$\textdegree.

We visualize these engineered shapes by placing the flashlight and the camera at, respectively, $d_f=(0.94 \pm 0.01)\,$m  and $d_c=(1.02 \pm 0.03)\,$m  above the plastic surface. To see the hologram movement, we slightly move the camera along a horizontal plane, keeping it at a distance $d_c$ above the surface.

Thus, we see that the IPT problems encourage students to experiment and find the relevant parameters first; later, students can strengthen their knowledge through more advanced techniques and topics. Initially, we performed various experiments to observe the patterns and propose the methods described in Section \ref{experiments}. Then, we were able to research different approaches to explain it, arriving at the Defocusing Microscopy technique. Additionally, we could expand our knowledge on optics simulations using Python and on holograms.

\section{Results and discussions}\label{results}
\subsection{Observed patterns}\label{a}

From our experiments with the various glass plates, we determine the conditions to form the main patterns, namely, (1) randomly scratched surfaces produce the halos and (2)~bright spots tend to generate diverging rays, which are optical distortions. 

\begin{figure}[hb]
    \centering
\includegraphics[scale=0.14]{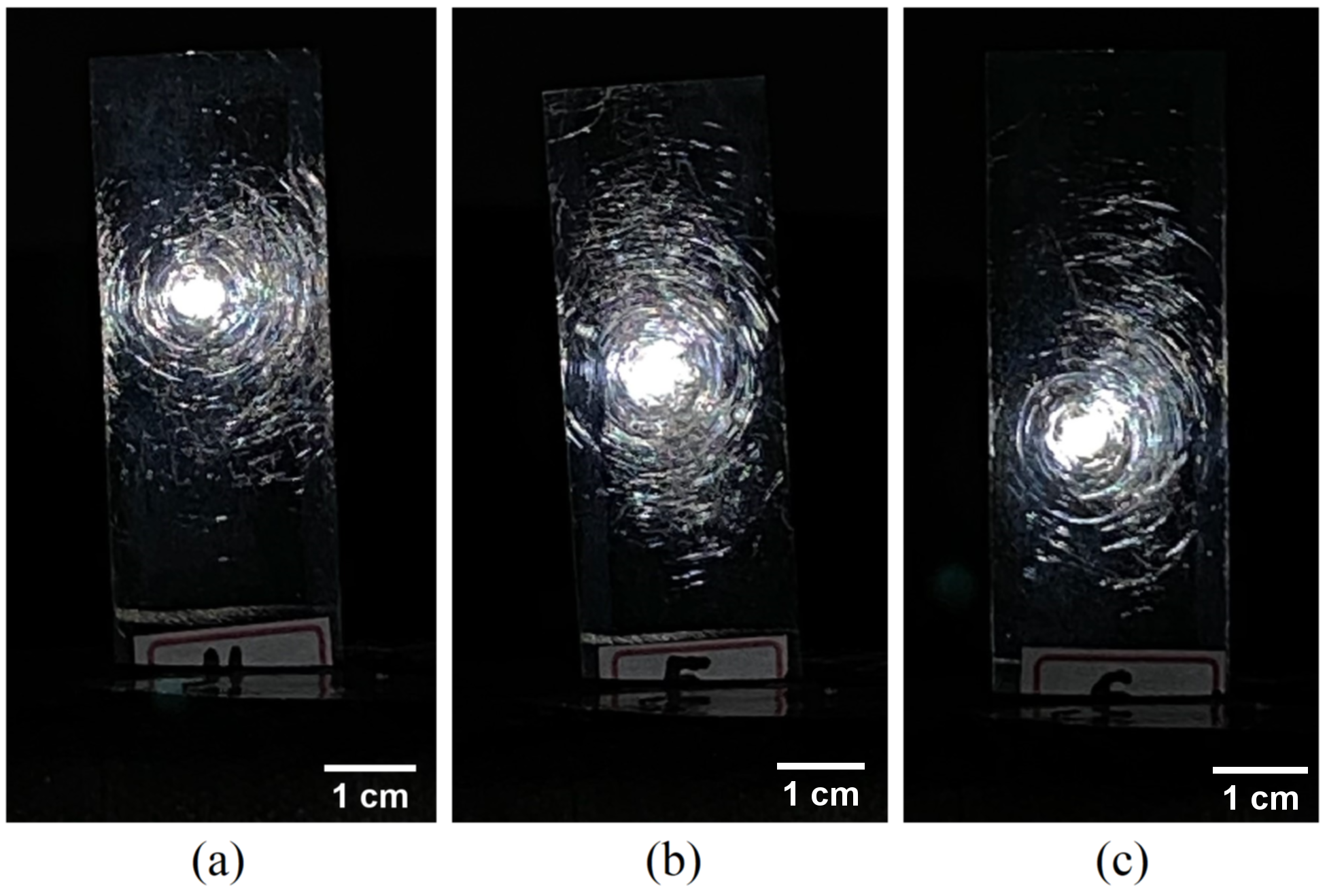}
    \caption{Halo formation for plates scratched with (a) a coarse grit, (b) with a medium grit and (c) with a fine grit.}
    \label{Fig. 4}
\end{figure}

First, the randomly scratched glass plates generate the glass halo, as shown in Fig.~\ref{Fig. 4} for plates 4, 5, and 6 described in Table~\ref{Table 1}. The halos are visible for all the randomly scratched plates regardless of the sandpaper used. As expected, the halo lines appear finer as we increase the sandpaper grit, while a coarse sandpaper produces a rougher and undefined circular pattern. We also observed the formation of this halo in other materials with randomly scraped surfaces, such as glass windows, cellphone screens, cars, spoons, and reflective surfaces in general (Fig. \ref{fig introducao}). This type of surface reflects light from all directions and a circular design is anticipated. 

\begin{figure}[htb]
    \centering
    \includegraphics[scale=0.12]{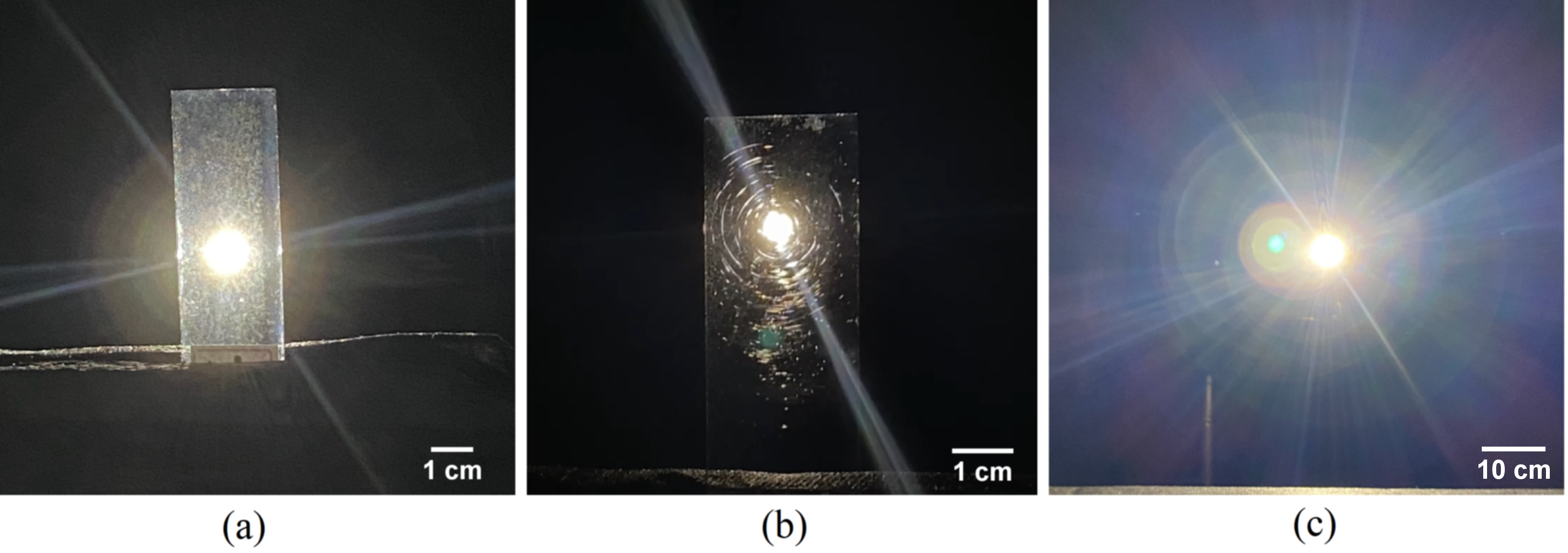}
    \caption{Distortion lines seen (a) through a polished glass (plate 1), (b) reflected on a scratched plate (plate 6) and (c) directly from the flashlight. The green dot also indicates that it is an imaging distortion.}
    \label{Fig. 5}
\end{figure}

The second main pattern is the formation of diverging lines, as seen in Fig.~\ref{Fig. 5}. They are an optical distortion in the imaging process and it can occur to any of the plates of Table~\ref{Table 1}, since it is related to the imaging process \cite{dai2022flare7k}. Indeed, in our experiments, we noticed that the lines are not visible to the bare eyes and that they rotate as the camera rotates (but not when the flashlight or the glass plate rotates). Researching the available literature along with experimentally checking the known results are a crucial part of tackling an IPT problem. In particular, this distortion is well-established in photography \cite{dai2022flare7k} and we have probed its behavior in our experimental setup, e.g. by rotating the experimental components, to check if we were really dealing with a distortion caused in the imaging process.

Lastly, when analyzing the reflection and refraction from polished plates, we noticed a saturated light circle surrounded by a blurred pattern that covered the entire glass surface - see now Fig.~\ref{Fig. 5}a. This blur is visible to the eyes and differs from the halos because it is not composed of scratches. 

\subsection{Scratch reconstruction}\label{reconstrução}

The halo pattern is just seen on the plates that were scratched with sandpaper, implying that the structures that generate this phenomenon are created in the process of scratching. At first, it is reasonable to assume that the scratch width could have any size, ranging from much smaller to much bigger than the visible light wavelength. However, the halo effect and scratches can be seen with the naked eye, which indicates that scratches with widths bigger than the visible light wavelength do exist. Most importantly, it can be seen during the experiments that parts of these visible scratches ``shine” to help create the halo effect, as seen in Fig. \ref{Fig. 4} for example, which indicates that these larger scratches must be taken into account to describe this phenomenon. It does not mean that other smaller structures do not exist: the ones with similar size, but larger than the wavelength would create diffraction or interference patterns, but these patterns or fringes are not observed in our microscopy experiment, while smaller structures would generate evanescent waves that would not propagate to an observer. Therefore, our object of interest are the structures with a width greater than the visible light wavelength.

\begin{figure}[ht]
    \centering
    \includegraphics[scale=0.35]{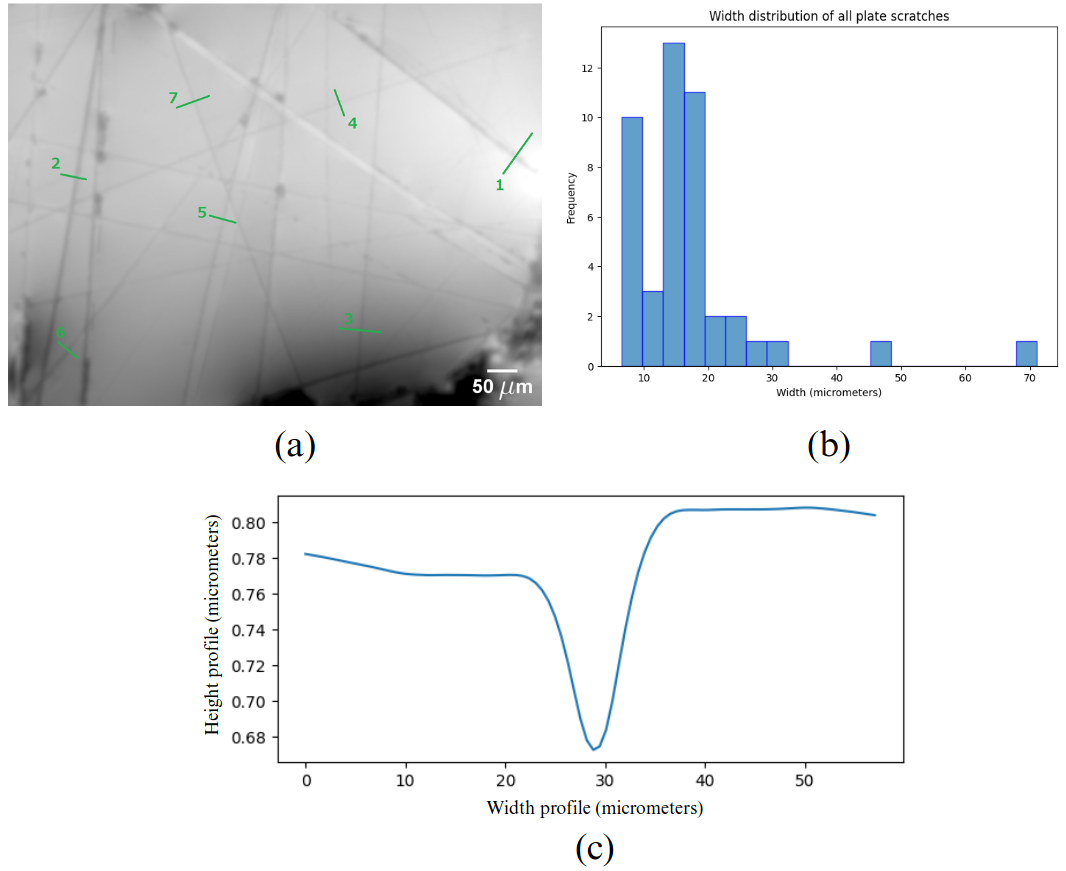}
    \caption{(a) Example of reconstruction of one of the plates (plate 5). The marked lines were used to calculate the mean width and depth. (b) Width distribution of all the 45 scratch samples collected in plates 4, 5 and 6. (c) Example of a lateral profile of a selected scratch.}
    \label{Fig. 6}
\end{figure}

Using the Defocusing Microscopy technique, we were able to reconstruct the topology of parts of the randomly scratched plates (4, 5 and 6) and measure the mean width and depth of the scratches on a plate, see Fig.~\ref{Fig. 6} (a). We also calculated the standard deviation of each quantity to measure how much the size distribution varies in each plate, see now Fig.~\ref{Fig. 6} (c). 

Taking in account the size of the structures and that they do not present a diffraction or interference pattern we assumed a geometric optics approach in the simulation to replicate the phenomena.

\begin{table}[h]
    \centering
    \renewcommand{\arraystretch}{1.3}
    \begin{tabular}{|c|c|c|c|c|}
        \hline
        Plate (Nº of samples) & Mean width & Width SD \\ \hline
        Plate 4 (15) & $18\mu$m & $10\mu$m  \\ \hline
        Plate 5 (14) & $14\mu$m & $3 \mu$m  \\ \hline
        Plate 6 (16) & $17\mu$m & $15\mu$m \\ \hline
    \end{tabular}
    \caption{Mean width and depth of the reconstructed scratches. Data available at \cite{sampaio_2024_14718426}.}
    \label{Table 3}
\end{table}

\subsection{Simulation}\label{d}

As detailed in Sec.~\ref{sec_simulation}, the final images created in our simulation are the patterns generated when light is shone over the scratched plates [given by points that obey Eq.~(\ref{cosdif})]. Thus, our simulation provided images of the pattern designs predicted using our theoretical model, and we compared them to our experimental results. 

\begin{figure}[ht]
    \centering
    \includegraphics[scale=0.4]{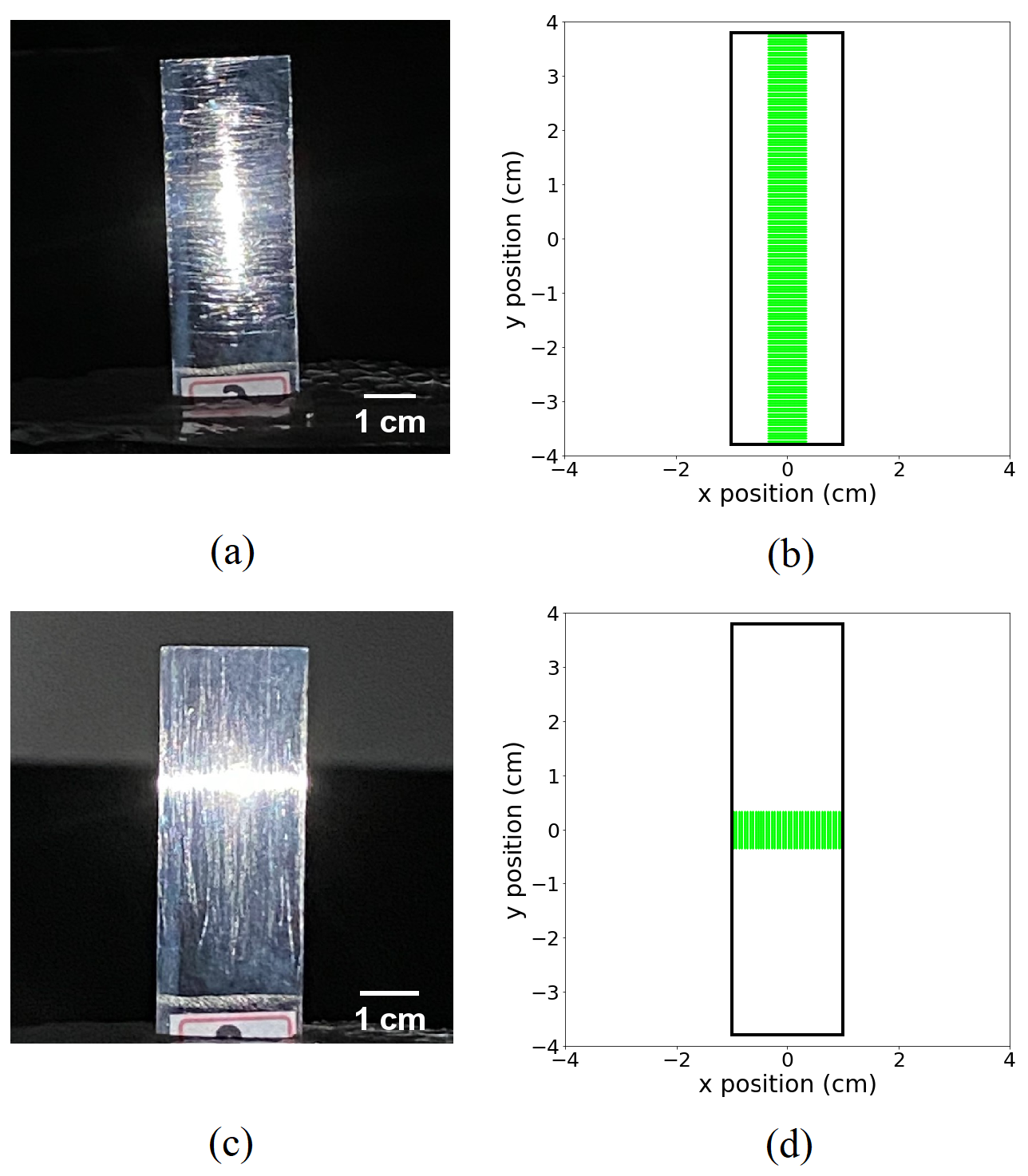}
    \caption{Comparison between (a) experimental and (b) computational results for a horizontally scratched plate, showing a vertical highlight. Similar analysis comparing (c) experimental and (d) computational patterns for a vertically scratched plate, showing a horizontal line. Both simulation trials we conducted with $d_f=d_c=0.55\,$m and a fixed number of scratches $N=400$.}
    \label{Fig. 7}
\end{figure}

Starting with the horizontally and vertically scratched glass (plates 2 and 3), in Fig.~\ref{Fig. 7} we show a comparison between experimental and theoretical results. In the results, we notice that a horizontally scraped plate highlights a vertical light pattern, and vertically scratched plates produce a horizontal highlight. We see a close similarity between the predicted and the experimental patterns.

\begin{figure}
    \centering
    \includegraphics[scale=0.45]{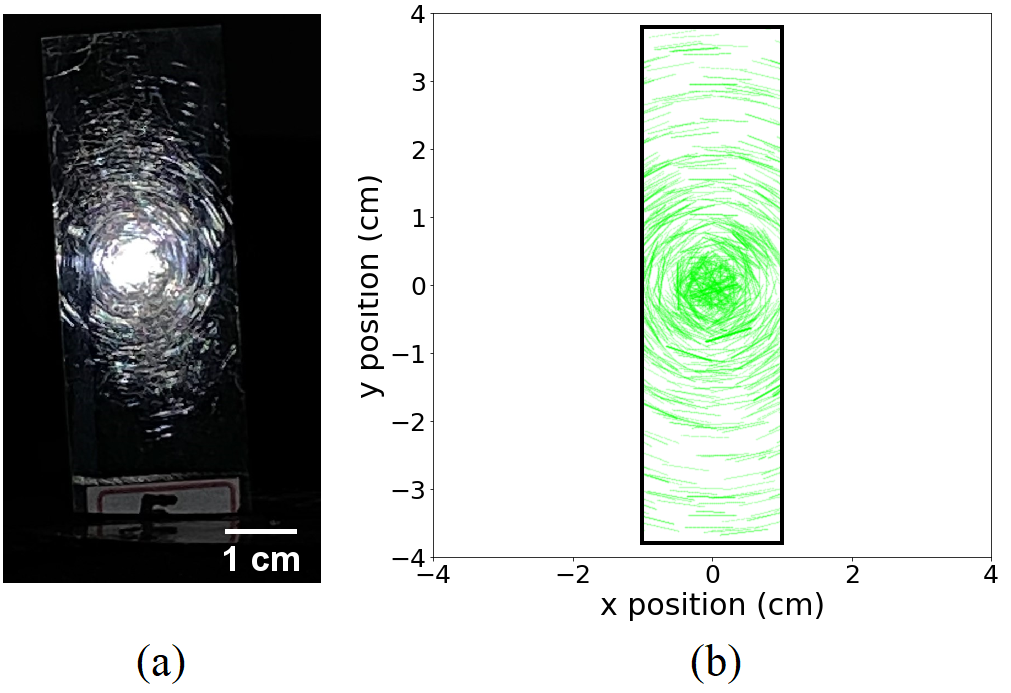}
    \caption{Comparison between (a) experimental and (b) simulation results for the halo at $d_f=d_c=(0.55\pm 0.01)\,$m. In the simulation, the number of scratches was $N=3000$.}
    \label{Fig. 8}
\end{figure}

Proceeding to the case of randomly scratched plates, we first compare the simulated and experimental results in Fig.~\ref{Fig. 8}, confirming that our model describes our results well. The consistency between the results indicate that the simplifications of our model are valid, and we can now use it to study other aspects of the phenomenon. The simulation becomes particularly useful when we lack the precision to perform a quantitative analysis - see Sec. \ref{proximos passos}. Thus, we studied the halo density by altering two factors in the simulation: the number of scrapes and the distance to the light-camera combination.

\begin{figure}
    \centering
    \includegraphics[scale=0.45]{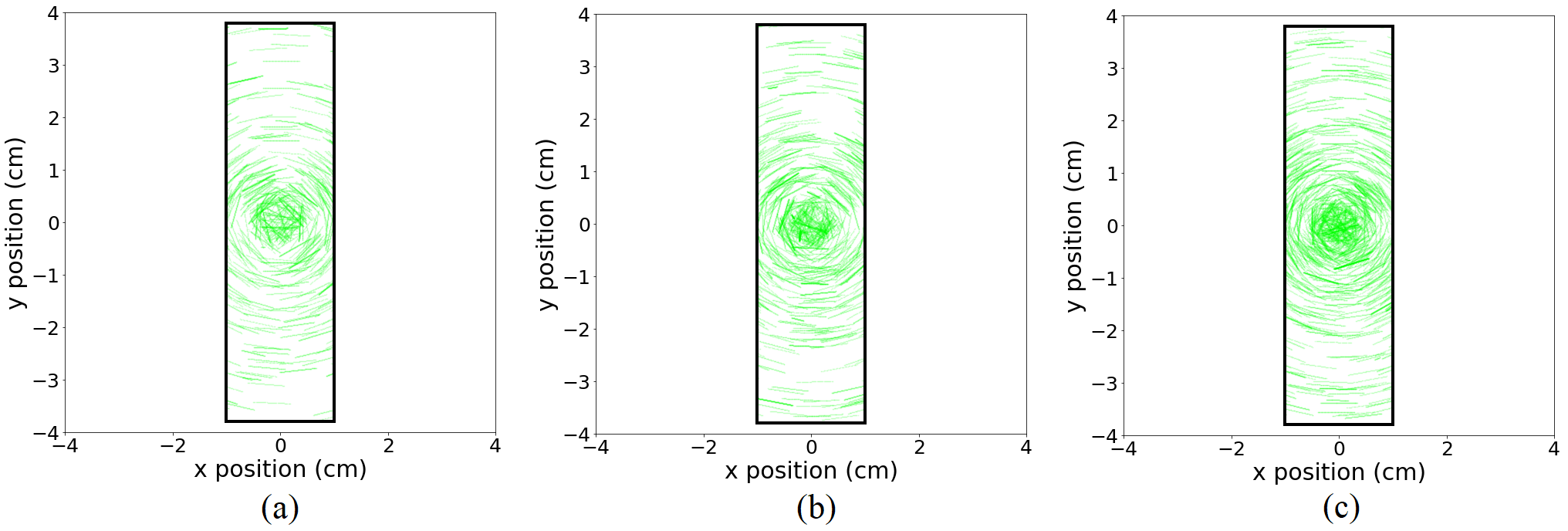}
    \caption{Halo result as we increase the number of scratches from (a) $N=1500$ to (b) $N=2000$ and, finally, to (c) $N=3000$. Trials conducted with a fixed distance of $d_f=d_c=0.55\,$m.}
    \label{Fig. 9}
\end{figure}

As we progressively increase the number of scratches on the surface, the halo density also increases, as we can observe in Fig.~\ref{Fig. 9}. This happens because for a larger number of scratches, more points obey the law of reflection for the small tolerance imposed, and the halo will appear fuller. 

\begin{figure}
    \centering
    \includegraphics[scale=0.45]{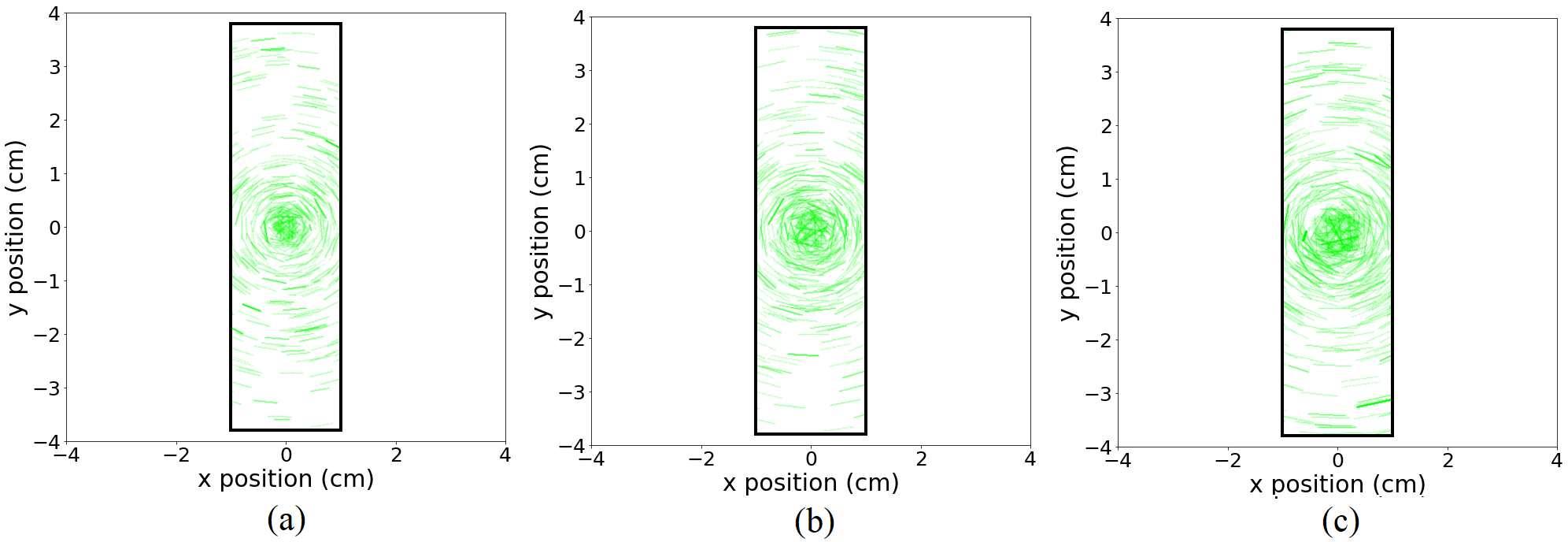}
    \caption{Halo formation for (a) $d_f=d_c=0.35\,$m, (b) $d_f=d_c=0.45\,$m and (c)~$d_f=d_c=0.55\,$m. In all these trials the number of scratches was constant ($N=2000$).}
    \label{Fig. 10}
\end{figure}

Similarly, for a fixed number of scratches, we see a fuller circle of halos when the plate is relatively far from the light-camera group - see now Fig.~\ref{Fig. 10}. As we move the surface closer to the light and the camera, fewer rays contact the scrapes and are reflected within the desired angle of the observer. These results are intuitive and confirm than our model can reproduce the phenomenon.

\subsection{Engineered patterns}\label{e}

\begin{figure}
    \centering
    \includegraphics[scale=0.46]{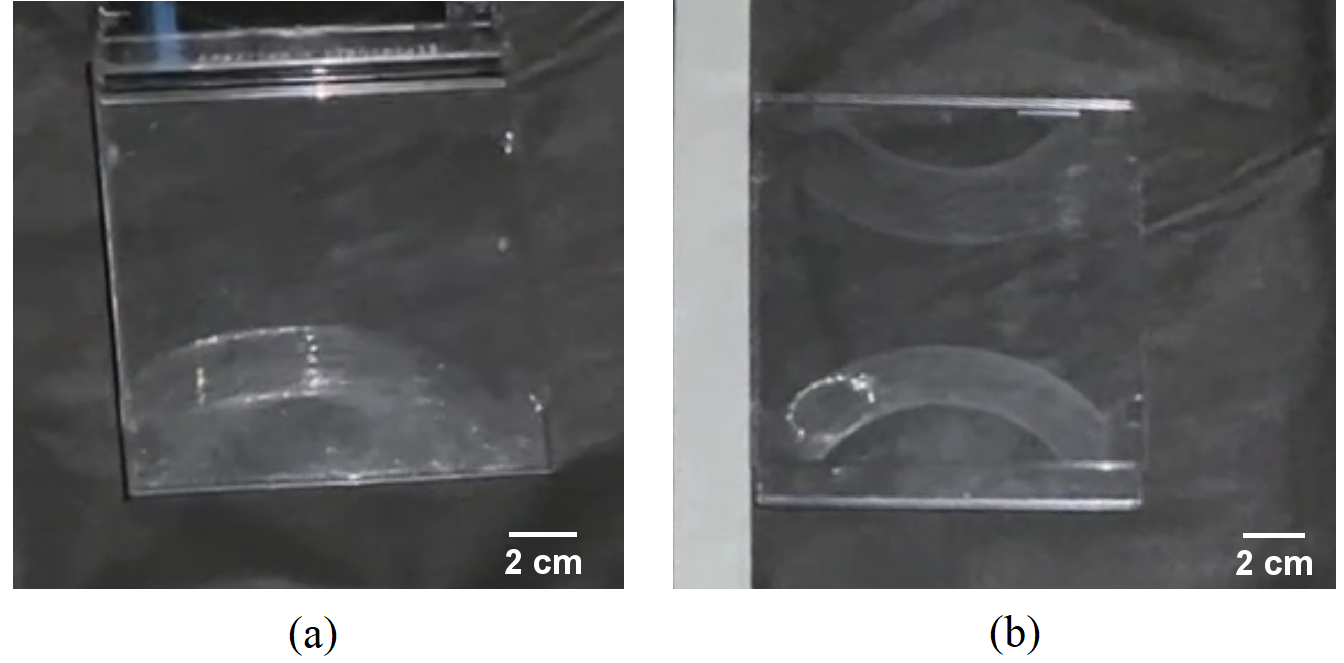}
    \caption{Two patterns created by scratching the CD cases: (a) a square hologram and (b) a circle hologram. The movement and 3D perspective is better seen in the available video.}
    \label{Fig. 11}
\end{figure}

We were able to engineer the two desired patterns, a square and a circle, using the specular holography method (see Fig.~\ref{Fig. 11}); the video showing the complete process for hologram creation and movement is available as supplementary material. The interesting feature of this method is that we can engineer any 3D shape or pattern by creating only circular scratches on the surface, provided that we follow the desired shape on a 2D blueprint.

We notice that the glass halo phenomenon does not form a 3D image. In the literature, the scratch hologram formation seems to demand scratches in a specific arch form, which act as curved spherical mirrors, producing either a real or virtual image \cite{inproceedings,plummer1992mechanically}, while the halo was formed with random locally linear scratches.

\subsection{Further studies}\label{proximos passos}

One way to improve this study is by verifying the equivalency between reflected and transmitted patterns. We used this fact to analyze only the reflected one in the simulation, but to certify it in the experiments one would need to capture the halo at the same position and angle on both sides of the surface, needing a very precise alignment mechanism.

It is extremely difficult to obtain quantitative analysis for the halo density. First, =we were not able to experimentally count the number of scratches to later compare the densities. Also, the point of light incidence on the glass needs to be accurate every time we decrease the distance, ensuring that it is the only varying factor. While we tried to maintain this caution in the experiments, one would need a more precise experimental setup to ensure the results and compare the halo density. Thus, we validated our simulation and use it to perform these analysis.

Lastly, an advancement can occur by obtaining a satisfactory result from the specular holography method on glass surfaces, rather than on a plastic material. The difficulty relies on scratching the glass in a controlled circular manner. To do that, one would need the appropriate pointer, e.g., a silicon carbide tool. An attempt to create a hologram on glass using laser engraving, as well as its limitations, was reported by Augier \cite{augier2011hologravure}.

\subsection{Pedagogical application}\label{aplicação pedagogica}

Our goal so far was to present the physics inside the IPT competition. We have shown the steps taken as we tackled the problem and our solution. 

Now we discuss its didactic use. First, the experimentation with scratched glass plates is easy and can be conducted to teach the principles of geometric optics since the scratches can reflect light in all directions, causing the circular halo pattern. This particular pattern can be seen in various objects of everyday life, as seen in Fig. \ref{fig introducao}, and can be easily reproduced. 

Also, the holography technique explored in this article shows how we can use simple tools to design optical patterns. In this context, the curved scratches act as spherical mirror, forming real or virtual images that we perceive as 3D images. Thus, these simple experiments communicate scientific knowledge related to reflection and image formation, being suitable for scientific fairs and didactic demonstrations in class.

Lastly, our simulation can act as a first contact with computational methods in physics for high school and undergraduate students. The simulation, written in Python, is accessible and exemplifies the use of arrays, functions and graphs. After understanding the principle of geometric optics, the student can use the simulation to analyze the role of each parameter in the optical outcome while learning a programming language.

\section{Conclusions}\label{conclusao}

We investigated the formation of optical patterns on glass. We observed experimentally that randomly scratched glass plates produce the glass halo phenomenon, highlighting the structural inhomogeneities. The second pattern we studied was the formation of diverging lines, which comes from a known optical distortion in the imaging process; indeed we observed that they rotate as we rotate the camera. We assumed that the phenomenon of light passing through a glass that has been mechanically scratched can be acceptably described using ray optics; indeed a simulation based on this assumption well reproduced our experiments. According to our results, the halo formation on everyday reflective objects, such as glass, is governed by specular reflection on various points along the scratches. We used our simulation to study the effect of two factors on the halo density, namely, (1) the number of scratches and (2) the distance between the surface and the camera-light source. As we progressively increase the number of scratches on the surface, the halo density also increases; similarly, we see a fuller circle of halos when the plate is relatively far from the light-camera position. Our methods can be used as a simple experimental demonstration of optical concepts in classes and science communication activities.

\section*{Acknowledgements}
The authors would like to thank Ariel Guimarães and Maria Vargas for their contribution, especially in the first stages of research. We also thank João Pedro Cunha for his help during experiments, and Anna Luisa Lemos for producing the supplementary video. MCOA and UA acknowledge financial support from CNPq, CAPES, and FAPEMIG.

\appendix
\section*{Appendix}
\setcounter{section}{1}

\subsection{Simulation} \label{appendix_simulation}
\begin{figure}
    \centering
    \includegraphics[width=0.3\linewidth, angle=-90]{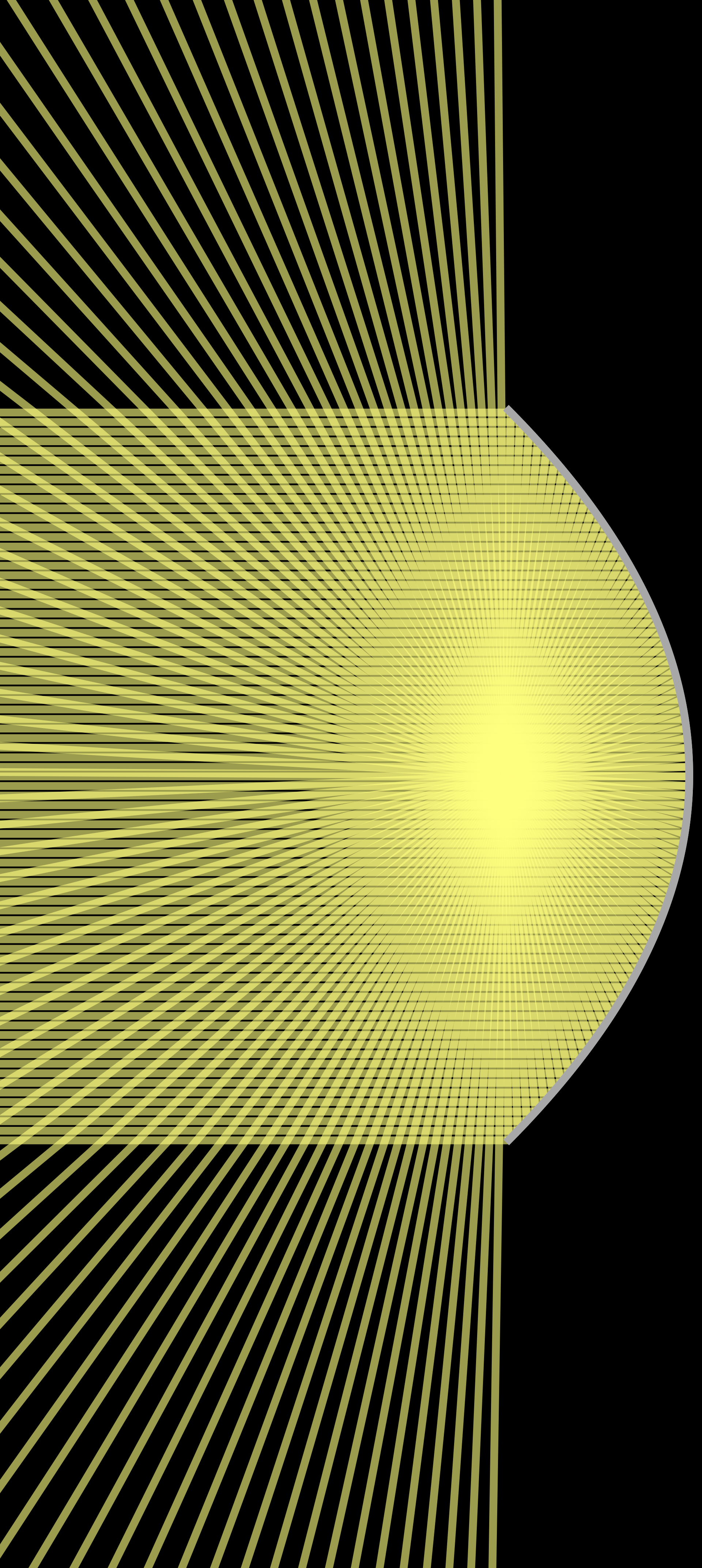}
    \caption{Schematic representation of the assumed scratch cross-sectional shape. The cross section of the scratch is assumed to be one such that rays from the light source (coming from above in the image) can reflect at any angle. The shape used to make the above image was a shallow parabolic mirror obtained using \cite{tu_2024_10547599}.}
    \label{fig:scratchCrossSection}
\end{figure}

In order to simulate the phenomenon, we follow \cite{li2017general}: suppose we have a reflective surface with a scratch whose curve can be parametrized by $\vec{r}(t)$ and whose unit tangent vector is $\hat{q}(t)$. If we have an arrangement such as in Fig.~\ref{Fig. 2}, then a point in the scratch will be seen if the following equation is true:

\begin{equation}
    \hat{E}(t)\cdot\hat{q}(t)=\hat{R}(t)\cdot\hat{q}(t), \label{lawreflection2}
\end{equation}
where $\vec{E}(t) = \vec{r}(t)-\vec{L}$, $\vec{R}(t) = \vec{C}-\vec{r}(t)$, and $\hat{E}(t)$ and $\hat{R}(t)$ represent their respective normalized vectors. One key assumption that has to be made is that the scratches have a cross sectional shape that is able to reflect at any angle, e.g. a shallow parabola or a shallow semicircle as seen in Fig.~\ref{fig:scratchCrossSection}; if this was not assumed, then additional conditions would need to be established in order to guarantee that the camera is inside of the region where the light reflected off the scratch could reach it. Another assumption needed is that the light source is far away and point-like, to guarantee that the light rays are parallel and reflect off the scratch as seen in Fig.~\ref{fig:scratchCrossSection}.

For a generic scratch curve, equation~(\ref{lawreflection2}) is not able to be solved analytically, and therefore needs root finding algorithms that naturally introduce a tolerance for the equation. Numerically, what the computer does is finding values of $t$ that satisfy:
\begin{equation}
    \left|\hat{E}(t)\cdot\hat{q}(t)-\hat{R}(t)\cdot\hat{q}(t)\right|\leq TOL, \label{unfinishedTolerance}
\end{equation}
or, calling $\theta_1$ the angle between $\hat{E}(t)$ and $\hat{q}(t)$ and, similarly, $\theta_2$ the angle between $\hat{R}(t)$ and $\hat{q}(t)$, the inequality in (\ref{unfinishedTolerance}) can be written as
\begin{equation}
    \left||\hat{E}(t)||\hat{q}(t)|cos(\theta_1)-|\hat{R}(t)||\hat{q}(t)|cos(\theta_2)\right|\leq TOL.
\end{equation}
Since $|\hat{E}(t)|=1$, $|\hat{R}(t)|=1$ and $|\hat{q}(t)|=1$, the camera or our eyes will capture the reflected beam if the following relation is satisfied, within a narrow tolerance,

\begin{equation}
    cos(\theta_1)-cos(\theta_2)\leq TOL. \label{cosdif}
\end{equation}

To illustrate the usage of equation (\ref{lawreflection}), we consider a the simplest example that can be solved analytically: a line. A line $\vec{r}(t)$ can be written as
\begin{equation}
\vec{r}(t) = \vec{P}_o + \hat{V}t,
\end{equation}
where $\vec{P}_o$ is the ``origin" of the line, i.e. the point where $t=0$, and $\hat{V}$ is the direction in which the line points with increasing $t$, which is also the tangent vector to the line. We will, without loss of generality, assume that $\hat{V}$ is normalized as notation implies.

Let us now consider the light source and the camera to be at symmetrical positions with respect to the $yz$ plane, that is, if $\vec{C}=(d,0,h)$, then $\vec{L}=(-d,0,h)$. With these vectors defined, we can use equation (\ref{lawreflection}) to get
\begin{equation}
\vec{r}(t)\cdot\hat{V} = \left(\frac{\vec{C}+\vec{L}}{2}\right)\cdot\hat{V}.
\end{equation}
As $\vec{C}$ and $\vec{L}$ are symmetrical, then $(\vec{C}+\vec{L})/2 = (0,0,h)$. Furthermore, since $\hat{V}$ is a vector parallel to the $xy$ plane, its $z$ component is zero, which makes the right hand side of the equation equal to zero. We now expand the left hand side using the definition of $\vec{r}(t)$:
\begin{equation}
\left(\vec{P}_o+\hat{V}t\right)\cdot\hat{V}=0.
\end{equation}
Therefore, the value of $t$ that satisfies this equation is
\begin{equation}
t = -\vec{P}_o\cdot\hat{V},
\end{equation}
and the point along the line that would be seen because of the reflection would be
\begin{equation}
\vec{r}(t)=\vec{P}_o-\left(\vec{P}_o\cdot\hat{V}\right)\hat{V}.
\end{equation}

For example, if the line was such that it passes through the point $\vec{P}_o=(1\mbox{\,m}, 0, 0)$ and has a slope of $45^\circ$ ($\hat{V}=1/\sqrt{2}(1\mbox{ m},1\mbox{ m},0)$) with respect to the $x$ axis, then the point on this line that would be seen is $1/2(1\mbox{ m},-1\mbox{ m})$.

One benefit of the simulation is that it ``automatically" accounts for the practical aspects of our retinas or the cameras' detectors. In practice, we observe not singular points, but certain segments that are illuminated from each scratch; and that is because light rays that are close enough to the point calculated to be the position of the glint also manage to reach the eye or camera because of the non-zero size of their apertures. Therefore, even for linear scratches it is better to simulate the phenomenon using equation (\ref{cosdif}), with a tolerance, rather than using the analytical solution obtained.

\subsection{Scratch reconstruction}\label{a2}

\begin{figure}
    \centering
    \includegraphics[scale=0.43]{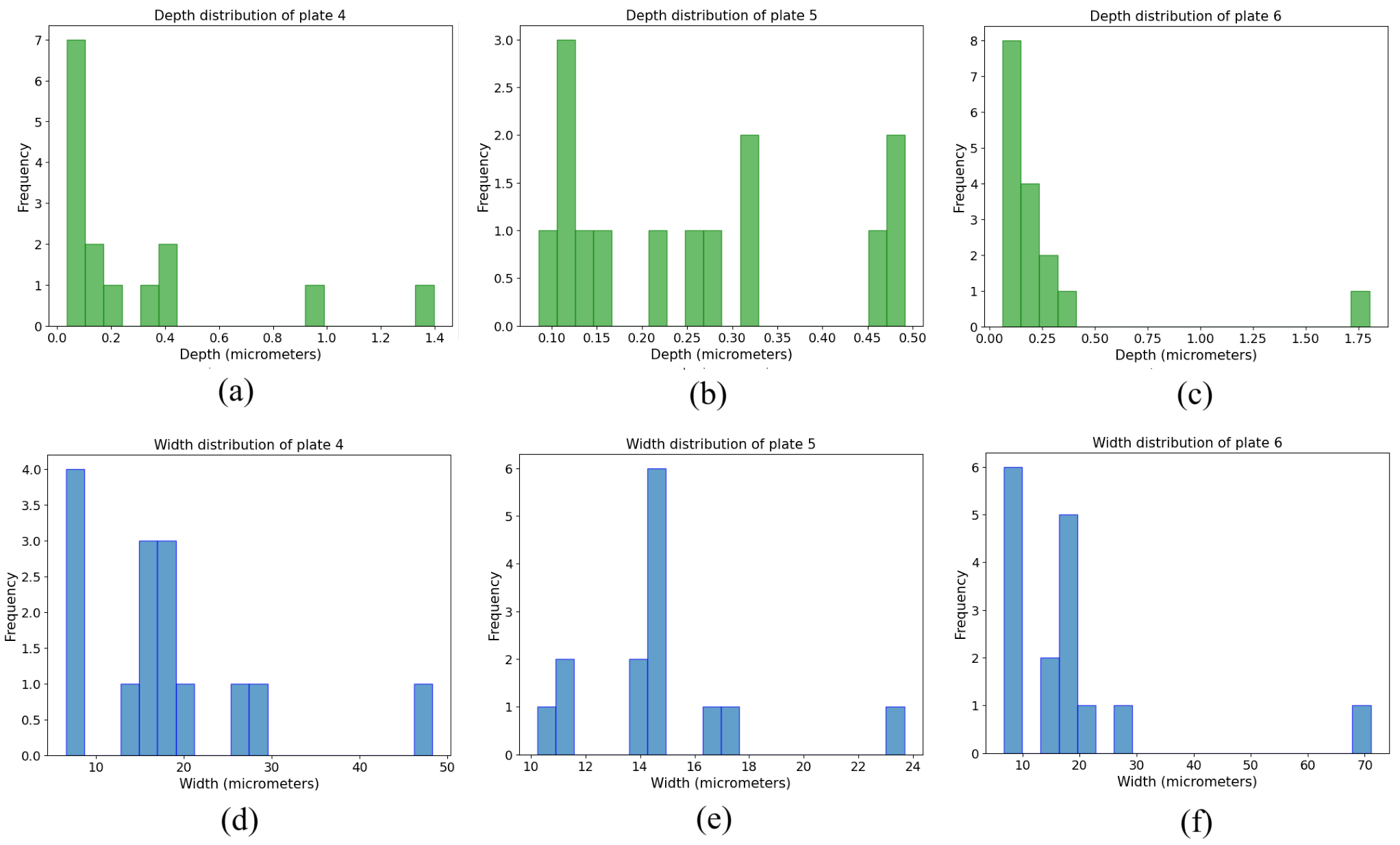}
    \caption{Distribution of the width and depth size for the randomly scratched plates.}
    \label{apendix fig2}
\end{figure}

As explained in Sec. \ref{reconstrução}, the reconstruction using the Defocusing Microscopy technique reveals how much the scratches vary in size; the standard deviation in Table \ref{Table 3} confirms that they are on the same scale (micrometers) but can differ significantly, since the sandpaper is not completely uniform and the grains can produce scratches of different sizes. We have shown the width distribution for all scratches used in our analysis (Fig. \ref{Fig. 6}). We highlight that they are larger than the wavelength of light and are concentrated around $18\, \mu$ m. Lastly, in Fig. \ref{apendix fig2} we present the distribution of the scratch width and depth for plates 4, 5 and 6.

\section*{References}
\bibliographystyle{vancouver}
\bibliography{references}

\begin{thebibliography}{10}

\bibitem{hahn2009light}
Hahn DW.
\newblock Light scattering theory.
\newblock Department of Mechanical and Aerospace Engineering, University of Florida. 2009:18.

\bibitem{elfouhaily2004critical}
Elfouhaily TM, Gu{\'e}rin CA.
\newblock A critical survey of approximate scattering wave theories from random rough surfaces.
\newblock Waves in random media. 2004;14(4):R1.

\bibitem{sylvain2005diffuse}
Sylvain M.
\newblock Diffuse reflection by rough surfaces: an introduction.
\newblock Comptes Rendus Physique. 2005;6(6):663-74.

\bibitem{maradudin2010light}
Maradudin AA.
\newblock Light scattering and nanoscale surface roughness.
\newblock Springer Science \& Business Media; 2010.

\bibitem{schroder2011modeling}
Schr{\"o}der S, Duparr{\'e} A, Coriand L, T{\"u}nnermann A, Penalver DH, Harvey JE.
\newblock Modeling of light scattering in different regimes of surface roughness.
\newblock Optics express. 2011;19(10):9820-35.

\bibitem{pinel2013electromagnetic}
Pinel N, Boulier C.
\newblock Electromagnetic wave scattering from random rough surfaces: Asymptotic models.
\newblock John Wiley \& Sons; 2013.

\bibitem{rice1951reflection}
Rice SO.
\newblock Reflection of electromagnetic waves from slightly rough surfaces.
\newblock Communications on pure and applied mathematics. 1951;4(2-3):351-78.

\bibitem{beckmann1987scattering}
Beckmann P, Spizzichino A.
\newblock The scattering of electromagnetic waves from rough surfaces.
\newblock Norwood. 1987.

\bibitem{warnick2001numerical}
Warnick KF, Chew WC.
\newblock Numerical simulation methods for rough surfacescattering.
\newblock Waves in random media. 2001;11(1):R1.

\bibitem{saillard2001rigorous}
Saillard M, Sentenac A.
\newblock Rigorous solutions for electromagnetic scattering from rough surfaces.
\newblock Waves in random media. 2001;11(3):103-37.

\bibitem{davies1954reflection}
Davies H.
\newblock The reflection of electromagnetic waves from a rough surface.
\newblock Proceedings of the IEE-Part IV: Institution Monographs. 1954;101(7):209-14.

\bibitem{bennett1961relation}
Bennett H, Porteus J.
\newblock Relation between surface roughness and specular reflectance at normal incidence.
\newblock JOSA. 1961;51(2):123-9.

\bibitem{croce1976light}
Croce P, Prod'homme L.
\newblock Light scattering investigation of the nature of polished glass surfaces.
\newblock Nouvelle Revue d'Optique. 1976;7(2):121.

\bibitem{ma2022implementation}
Ma Z, Wang H, Chen Q, Xue Y, Pan Y, Shen Y, et~al.
\newblock Implementation of empirical modified generalized Harvey--Shack scatter model on smooth surface.
\newblock Journal of the Optical Society of America B. 2022;39(7):1730-5.

\bibitem{toporkov2000numerical}
Toporkov JV, Brown GS.
\newblock Numerical simulations of scattering from time-varying, randomly rough surfaces.
\newblock IEEE Transactions on Geoscience and Remote Sensing. 2000;38(4):1616-25.

\bibitem{yan2022deep}
Yan K, Yang S, Zhao Y, Ma C, Jin Y, Wang S.
\newblock Deep learning for light scattering computation: Reconstructing light scattering fields from 1-D randomly rough surfaces as an example.
\newblock Computer Physics Communications. 2022;270:108183.

\bibitem{zhang2025generalized}
Zhang Y, Zhong Z, Zhang B.
\newblock Generalized Beckmann-Kirchhoff scattering model for ultra-smooth surfaces.
\newblock Journal of Modern Optics. 2025:1-12.

\bibitem{inproceedings}
Beaty W.
\newblock Drawing Holograms by Hand.
\newblock vol. 5005; 2003. .

\bibitem{duke2013drawing}
Duke T.
\newblock Drawing Light-fields: Hand-drawn Approaches to Abrasion Holography.
\newblock In: Journal of Physics: Conference Series. vol. 415. IOP Publishing; 2013. p. 012033.

\bibitem{li2017general}
Li B, Kang ZY, Lin J, Yeo Y, Tan G.
\newblock General theories of reflection and transmission scratch holograms.
\newblock Canadian Journal of Physics. 2017;95(5):432-9.

\bibitem{ipt}
IPT 2023 Problems; 2023.
\newblock Available from: \url{https://2023.iptnet.info/problems/}.

\bibitem{iptrules}
Official Rules; 2024.
\newblock Available from: \url{https://iptnet.info/official-rules/}.

\bibitem{sampaio_2024_10899678}
Sampaio M, do~Valle G, Kazazis I, Alvim RG, Zanni P, Roberto L, et~al.. I IPT Conference Proceedings. Zenodo; 2024.
\newblock Available from: \url{https://doi.org/10.5281/zenodo.10899678}.

\bibitem{agero2004defocusing}
Agero U, Mesquita L, Neves B, Gazzinelli R, Mesquita O.
\newblock Defocusing microscopy.
\newblock Microscopy research and technique. 2004;65(3):159-65.

\bibitem{roma2014total}
Roma P, Siman L, Amaral F, Agero U, Mesquita O.
\newblock Total three-dimensional imaging of phase objects using defocusing microscopy: Application to red blood cells.
\newblock Applied Physics Letters. 2014;104(25).

\bibitem{Lages:18}
Lages E, Cardoso W, Almeida GFB, Siman L, Mesquita O, Mendon\c{c}a CR, et~al.
\newblock Measurement of the refractive index profile of waveguides using defocusing microscopy.
\newblock Appl Opt. 2018 Oct;57(29):8699-704.
\newblock Available from: \url{https://opg.optica.org/ao/abstract.cfm?URI=ao-57-29-8699}.

\bibitem{sampaio_2024_14718426}
Sampaio MP, Alvim RG, Kalil FK, Aguiar MCdO, Agero U. Research data for the paper "Analysis of optical pattern formation on glass: a solution for an International Physicists' Tournament problem". Zenodo; 2024.
\newblock Available from: \url{https://doi.org/10.5281/zenodo.14718426}.

\bibitem{neoonlineWhatPlastic}
What plastic is a {C}{D} jewel case? --- neo-online.co.uk; 2023.
\newblock [Accessed 15-05-2024].
\newblock \url{https://neo-online.co.uk/what-plastic-is-a-cd-jewel-case/}.

\bibitem{dai2022flare7k}
Dai Y, Li C, Zhou S, Feng R, Loy CC.
\newblock Flare7k: A phenomenological nighttime flare removal dataset.
\newblock Advances in Neural Information Processing Systems. 2022;35:3926-37.

\bibitem{plummer1992mechanically}
Plummer WT, Gardner LR.
\newblock A mechanically generated hologram?
\newblock Applied optics. 1992;31(31):6585-8.

\bibitem{augier2011hologravure}
Augier {\'A}G, S{\'a}nchez RB.
\newblock Hologravure as a computer-generated and laser engraved scratch hologram.
\newblock Optics communications. 2011;284(1):112-7.

\bibitem{tu_2024_10547599}
Tu YT. Ray Optics Simulation. Zenodo; 2024.
\newblock Available from: \url{https://doi.org/10.5281/zenodo.10547599}.

\end{thebibliography}

\end{document}